\documentclass[twocolumn,showpacs,superscriptaddress,amssymb,10pt,prl,floatfix,aps,longbibliography]{revtex4-2}
\usepackage[dvips]{graphicx}

\usepackage{longtable}
\usepackage{dcolumn}
\usepackage[dvips]{graphicx}
\usepackage{bm}
\usepackage{lipsum}
\usepackage{bbm}
\usepackage{color}

\usepackage{times}
\usepackage{nicefrac}
\usepackage{amsmath}
\usepackage{amsfonts}
\usepackage{amssymb}
\usepackage{amsthm}
\usepackage[normalem]{ulem}
\newcolumntype{.}{D{x}{}{-1}}
\usepackage{orcidlink}
\usepackage{epsfig}
\usepackage{leftidx,amsmath}
\usepackage{natbib}
\usepackage[utf8]{inputenc}
\usepackage{braket}
\usepackage{makecell}
\usepackage[flushleft]{threeparttable}
\usepackage{multirow}
\usepackage{isotope}
\usepackage{array}
\usepackage{siunitx}

\begin{document}
%%%%%%%%%%%%%%%%%%%%%%%%%%%%%%%%%%%%%%%%%%%%%%%%%%%%%%%%%%%%%%%%%%%%%%%%%%%%%%%

\title{The two-loop Lamb shift: the final missing contribution}

\author{J.~Sommerfeldt, \orcidlink{0000-0002-3471-7494}}
\thanks{jonas.sommerfeldt@lkb.upmc.fr}
\affiliation{Max Planck Institute for Nuclear Physics, D-69117 Heidelberg, Germany}
\affiliation{Laboratoire Kastler Brossel, Sorbonne Universit\'e, CNRS, ENS-PSL Research University, Coll\`ege de France, Case\ 74;\ 4, place Jussieu, F-75005 Paris, France}

\author{V.~A.~Yerokhin, \orcidlink{0000-0002-2328-8444}}
\affiliation{Max Planck Institute for Nuclear Physics, D-69117 Heidelberg, Germany}

\author{Z.~Harman, \orcidlink{0009-0006-0554-1537}}
\affiliation{Max Planck Institute for Nuclear Physics, D-69117 Heidelberg, Germany}

\author{C.~H.~Keitel, \orcidlink{0000-0002-1984-1470}}
\affiliation{Max Planck Institute for Nuclear Physics, D-69117 Heidelberg, Germany}

\date{\today \\[0.3cm]}

\begin{abstract}

The last significant unknown contribution to the two-loop Lamb shift, namely the electron
self-energy with vacuum-polarization insertion into the photon line, is determined
including higher-order effects in the parameter $\alpha Z$,
with $\alpha$ being the fine-structure constant and $Z$ the nuclear charge number.
Our calculation
improves the accuracy of theoretical predictions for energy levels and transition energies in heavy hydrogen-like
and lithium-like ions by up to one order of magnitude, enabling more stringent comparisons between theory and available experimental results.
These results pave the way for more precise tests of quantum electrodynamics in the strong-field regime,
as well as for accurate determinations of nuclear charge radii.

\end{abstract}
\maketitle
\textit{Introduction---}
Few-electron highly charged ions (HCIs) combine the strong electric field of the nucleus with a relatively simple electronic structure, making them ideal systems for testing bound-state quantum electrodynamics (QED), searching for physics beyond the Standard Model, and determining nuclear properties~\cite{sbgk2018, kscs2018, ind2019}. The strength of the nuclear Coulomb field in HCIs is characterized by the parameter $\alpha Z$, where $\alpha \approx 1/137$ is the fine-structure constant and $Z$ the nuclear charge number. For very heavy HCIs
with $\alpha Z \approx 1$, theoretical calculations of atomic properties must be performed without expansion in $\alpha Z$, accounting for the Coulomb interaction between the bound electrons and the nucleus to all orders.

Theoretical calculations of QED effects to first order in $\alpha$ but to all orders in $\alpha Z$ had been established  by
the end of the 1980s \cite{mohr:74:a,soff:88:vp,manakov:89:zhetp}, enabling first stringent tests of bound-state QED in the regime
of strong electron-nucleus coupling \cite{mps1998}. The rapid development of experimental techniques, however, soon called for
theoretical calculations of effects of second order in $\alpha$ without expansion in $\alpha Z$.
The seminal 1991 experiment by Schweppe et al.~\cite{schweppe:91} on Li-like uranium, followed by numerous subsequent measurements
with enhanced precision
\cite{beiersdorfer:93,beiersdorfer:95,beiersdorfer:98,brandau:03,beiersdorfer:05},
triggered extensive theoretical efforts to develop new techniques
required for calculations of $\alpha^2$ effects to all orders in $\alpha Z$.

\begin{figure}[t]
    \centering
    \includegraphics[width=0.8\linewidth]{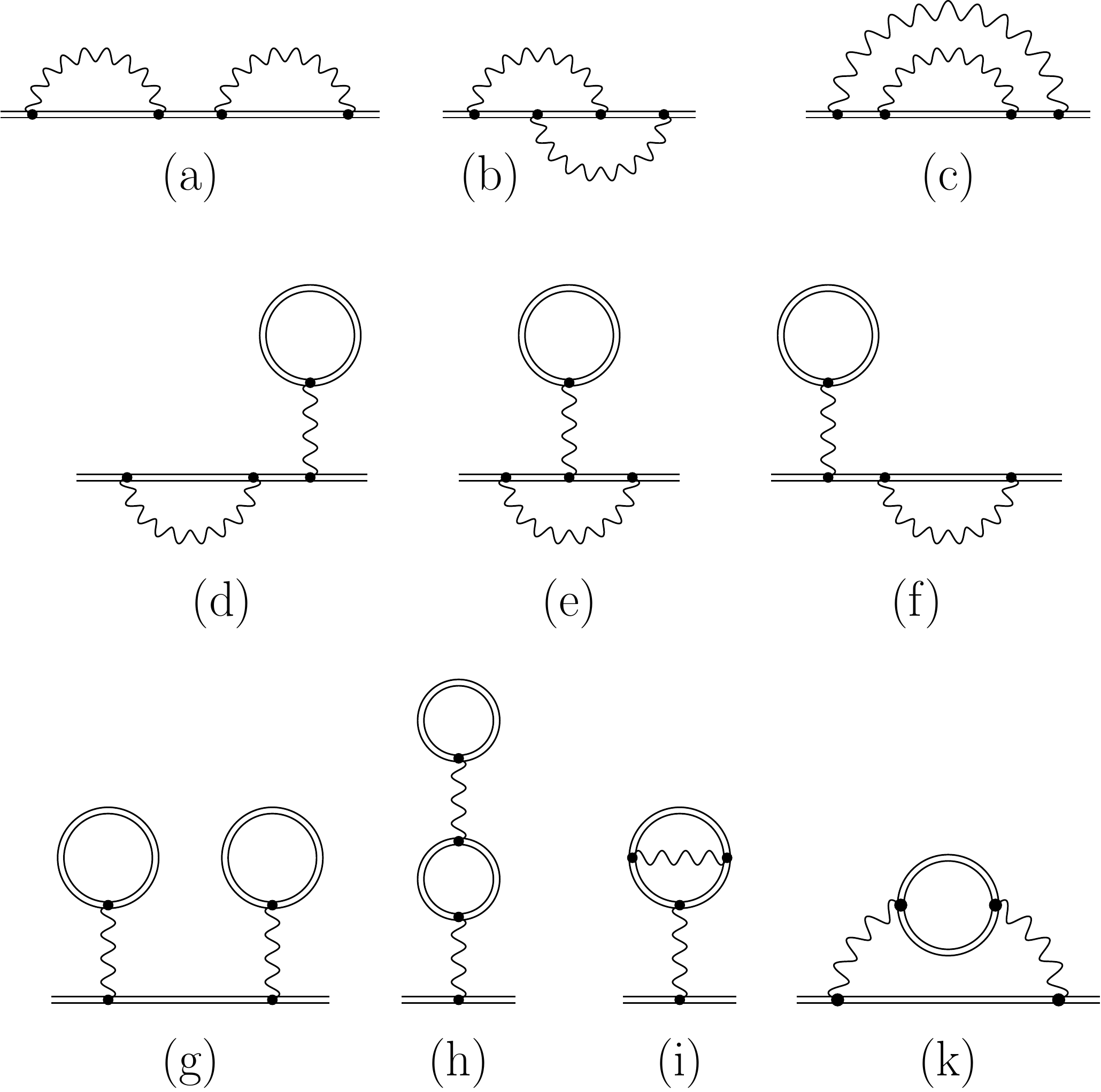}
    \caption{Feynman diagrams representing the two-loop Lamb shift. The double line denotes the electron in the Coulomb field of the nucleus and the wavy lines are virtual photons. The gauge invariant subsets of diagrams are the two-loop SE (SESE): (a)-(c), mixed VP and SE (SEVP): (d)-(f), two-loop VP (VPVP): (g)-(i), and SE with VP insertion (SVPE): (k).}
    \label{Fig:2LoopFeynman}
\end{figure}

Already by the end of the 1990s, most of the {\em two-electron} $\alpha^2$ effects, including the so-called
screened QED and two-photon exchange corrections, had been calculated
\cite{blundell:93:b,persson:96:2el,yerokhin:99:sescr,yerokhin:00:prl}.
By contrast, the {\em one-electron} $\alpha^2$ effects, referred to here as the two-loop Lamb shift (shown in Fig.~\ref{Fig:2LoopFeynman}),
proved to be a much more challenging problem. Although several individual contributions were calculated
with comparatively little difficulty
\cite{mitrushenkov:95,mas1996,persson:96:pra,bpgs1997,pbsp1998},
the theoretical prediction for the two-loop Lamb shift remained unavailable for a long time.
The main obstacle was the two-loop self-energy (SESE subset in Fig.~\ref{Fig:2LoopFeynman}),
which was expected to provide the dominant contribution.

The breakthrough in the SESE correction was achieved in a series of investigations \cite{yerokhin:03:prl,yerokhin:05:sese,yerokhin:06:prl,yerokhin:18:sese,yerokhin:25:sese} after years of dedicated efforts.
After that, most of remaining diagrams in Fig.~\ref{Fig:2LoopFeynman} had been calculated by 2008 \cite{yerokhin:08:twoloop}.
However, the long-standing project of calculating the two-loop Lamb shift has not been fully completed yet.
The reason is that the diagrams in Fig.~\ref{Fig:2LoopFeynman}(i)--(k) were calculated only in the so-called free-loop
approximation, in which the vacuum-polarization loop includes only the first nonvanishing term in the expansion in the
nuclear Coulomb field.
Although the uncertainty associated with these unfinished diagrams determined the overall theoretical accuracy in
hydrogen-like ions \cite{yerokhin:15:Hlike}, no further progress on the two-loop Lamb shift was made for a long time.
More recently, the same uncalculated corrections have begun to limit the theoretical precision also for helium-like
and lithium-like ions \cite{malyshev:23,yerokhin:25:lilike}.

A step towards completing the two-loop Lamb-shift calculations was recently made by Volkov et al.~\cite{vyhk2025}, who
computed the diagrams in Fig.~\ref{Fig:2LoopFeynman}(h) and (i). Although their calculation was not performed to all
orders in $\alpha Z$, the employed expansion proved to be very effective for a wide range of $Z$ up to uranium.
This eliminated one of the two remaining sources of uncertainty in the two-loop Lamb shift.
The goal of the present Letter is to perform a complete calculation of the SVPE correction shown in
Fig.~\ref{Fig:2LoopFeynman}(k), thereby eliminating the last remaining source of uncertainty and completing
the long-standing project of calculating the two-loop Lamb shift.

%Relativistic units $\hbar=c=m_e=1$ are used throughout this Letter unless stated otherwise.

\textit{Theoretical Framework---}
Using standard methods of bound-state QED \cite{shabaev:02:rep}, the energy shift due to the SVPE diagram shown in Fig.~\ref{Fig:2LoopFeynman}(k) is expressed as
\begin{widetext}
\begin{equation} \label{Eq:ESVPE}
\begin{aligned}
   E_\text{SVPE} = \frac{(4\pi\alpha)^2}{(2\pi i)^2} \int_{C_F}d\omega \int_{C_F}dz\int &d^3\boldsymbol{x}_1~d^3\boldsymbol{x}_2~ d^3\boldsymbol{x}_3~d^3\boldsymbol{x}_4~\psi^\dagger_a(\boldsymbol{x}_1)\alpha_\mu G\left(\boldsymbol{x}_1, \boldsymbol{x}_4, \epsilon_a-\omega\right)\alpha_\sigma\psi_a(\boldsymbol{x}_4) \\
   &\times D^{\mu\nu}(\boldsymbol{x}_{12},\omega) \text{Tr}\left[\alpha_\nu G\left(\boldsymbol{x}_2, \boldsymbol{x}_3, z+\frac{\omega}{2}\right) \alpha_\rho G\left(\boldsymbol{x}_3, \boldsymbol{x}_2, z-\frac{\omega}{2}\right)\right] D^{\rho\sigma}(\boldsymbol{x}_{34},\omega) ~,
\end{aligned}
\end{equation}
\end{widetext}
where $\psi_a$ is the reference-state electron wave function with the energy $\epsilon_a$, $G\left(\boldsymbol{x}_2, \boldsymbol{x}_1, z\right)$ is the Dirac-Coulomb Green's function, $D^{\mu\nu}(\boldsymbol{x}_{12},\omega)$ is the photon propagator,
$\alpha^{\mu} = (1,\boldsymbol{\alpha})$ are the Dirac matrices, and $C_F$ is the standard Feynman integration contour.

\begin{figure}
    \centering
    \includegraphics[width=1.0\linewidth]{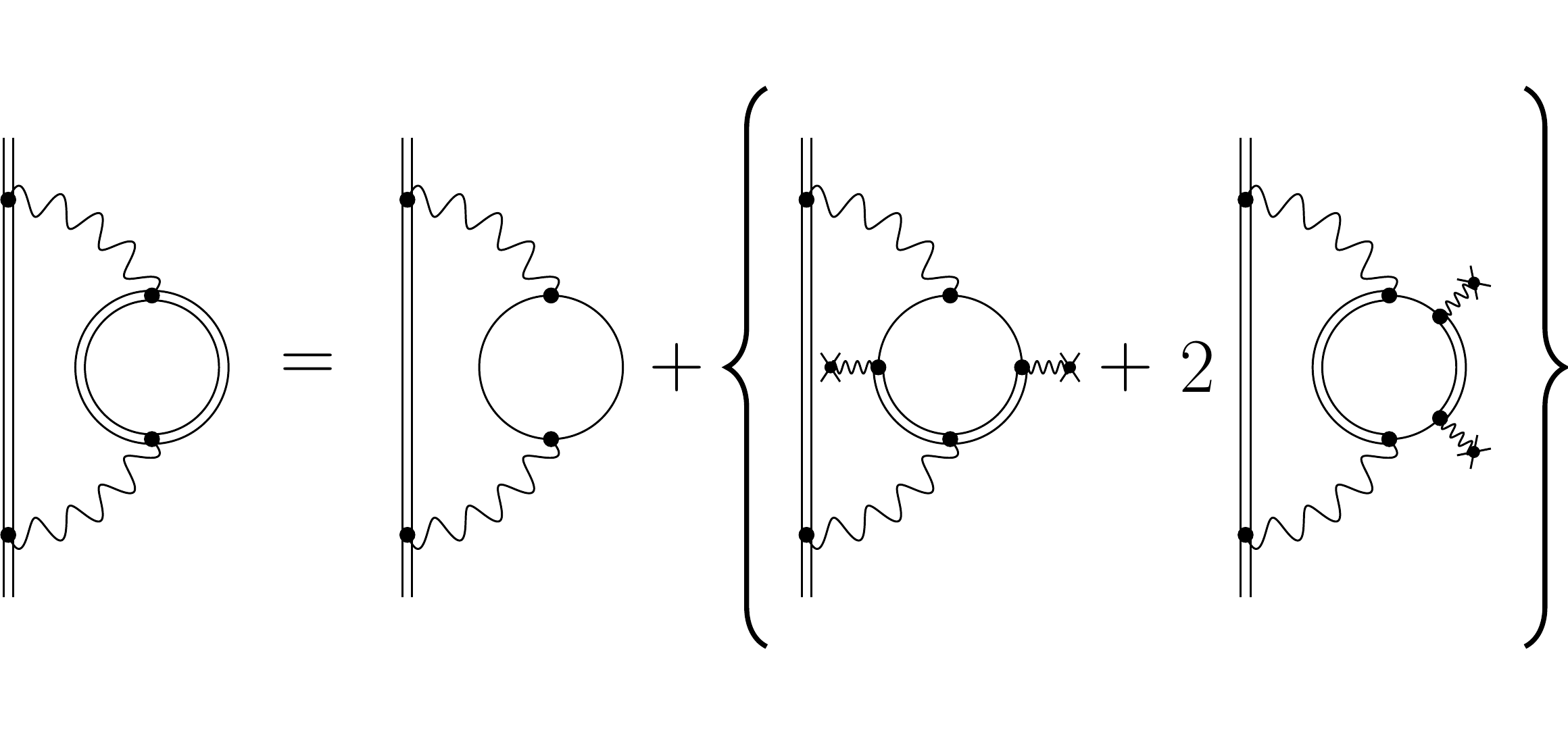}
    \caption{Separation of the SVPE correction into the free-loop part and the higher-order remainder.
    Single lines represent the free electron propagator and wavy lines terminated by crosses are interactions with the Coulomb field.
    Note that the electron loops with odd number of vertices vanish according to the Furry theorem~\cite{furr1937}.}
    \label{Fig:FeynmanExp}
\end{figure}

For the numerical evaluation, it is advantageous to split the SVPE energy shift into two parts, separating out the
free-electron vacuum-polarization loop contribution (see Fig.~\ref{Fig:FeynmanExp}),
\begin{equation}
    E_\text{SVPE} = E_\text{SVPE}^{(0)} + E_\text{SVPE}^{(2+)}~.
\end{equation}
Here $E_\text{SVPE}^{(0)}$ is the free-loop part, corresponding to the first diagram on the right side of Fig.~\ref{Fig:FeynmanExp}, and $E_\text{SVPE}^{(2+)}$ is the higher-order correction, displayed by the two diagrams in curly brackets in Fig.~\ref{Fig:FeynmanExp}.

The calculation of the free-loop contribution $E_\text{SVPE}^{(0)}$ was first performed in the 1990s \cite{pllp1996,mas1996} and is well established
at present \cite{yis2008}. This contribution can be related to the well-known one-loop self-energy (SE) correction,
in which the standard photon propagator is replaced by the ``radiatively dressed'' photon propagator, given (in $D=4-2\epsilon$ dimensions) by
\cite{mas1996,yis2008}
\begin{equation}
    \frac{1}{k^2+i\delta} \to -\frac{\alpha C_\epsilon}{4\pi}\int_0^1 dt~\frac{t^2(1-t^2/3)}{[m^2-k^2(1-t^2)/4-i\delta]^{1+\epsilon}}~
\end{equation}
where $C_\epsilon = (4\pi)^\epsilon \Gamma(1+\epsilon)$.
The general evaluation scheme is analogous to that for the one-loop SE correction \cite{yerokhin:99:pra}, with
the electron propagator in the self-energy loop expanded into the zero-, one-, and many-potential terms
according to the number of the Coulomb interactions.
The zero- and one-potential terms are evaluated in momentum space and their computation is very similar to the one-loop SE but more time-consuming due to the presence of additional integrations over Feynman parameters. The many-potential term is evaluated in coordinate space and is equivalent to the calculation of the one-loop SE with an effective photon mass and an additional numerical integration. For further details we refer the reader to the literature~\cite{mas1996,yis2008}.

The calculation of the higher-order SVPE contribution $E^{(2+)}_\text{SVPE}$  is the central part of this work.
This contribution is ultraviolet finite, if the gauge invariance is ensured at all stages of the computation.
It can be shown that spurious gauge-noninvariant terms are absent in the present treatment
based on the partial-wave expansion of electron propagators, similarly to previous analogous
calculations of the vacuum-polarization correction \cite{soff:88:vp,artemyev:99}.
The higher-order SVPE contribution is computed by taking a point-by-point difference of the complete SVPE
expression \eqref{Eq:ESVPE} and its free-loop limit.

Inserting the multipole expansions of the electron and photon propagators into Eq.~\eqref{Eq:ESVPE} and applying the standard angular-momentum algebra~\cite{jbs1988}, we obtain
\begin{equation} \label{Eq:AmpFinal}
    \begin{aligned}
         E&_\text{SVPE}^{(2+)} = \frac{(4\pi\alpha)^2}{(2\pi i)^2} \sum_{J_1 J_2,\kappa \kappa'\kappa''} \sum_{\mu \mu'\mu''}
        \\ & \times I_{J_1}(a,\kappa'\mu';\kappa\mu,\kappa''\mu'') I_{J_2}(\kappa''\mu'',\kappa\mu;\kappa'\mu',a)\\
        & \times\int_{C_F} d\omega~\int_{C_F} dz~R_{J_1}(a\kappa';\kappa\kappa'') R_{J_2}(\kappa''\kappa;\kappa' a)
        \nonumber \\ &
        - \mbox{\rm Subtractions}
        ~,
    \end{aligned}
\end{equation}
where $\kappa',\mu'$ and $\kappa'',\mu''$ are the Dirac angular-momentum quantum numbers and the angular-momentum projections of the two electron propagators in the fermion loop, $\kappa, \mu$ are the Dirac angular-momentum quantum number and the angular-momentum projection of the self-energy electron propagator, $J_1$ and $J_2$ are the angular-momentum quantum numbers of the two photon propagators,
and $R_J$ and $I_J$ are the radial and angular integrals introduced in Ref.~\cite{syhc2026}. The subtractions in Eq.~(\ref{Eq:AmpFinal}) correspond
to the removal of the free-loop part.

\begin{figure}
    \centering
    \includegraphics[width=0.9\linewidth]{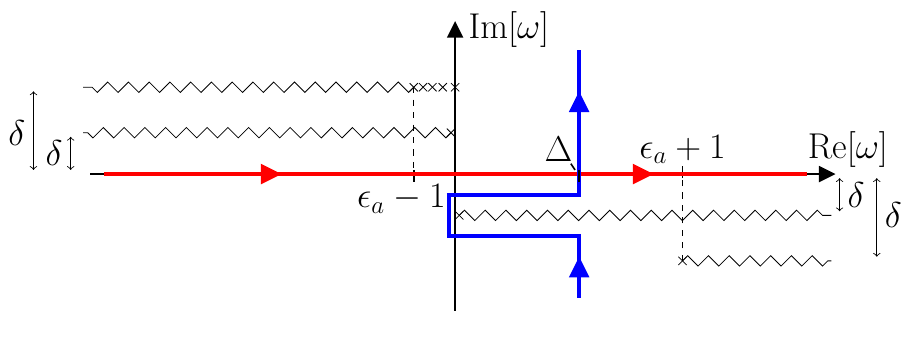}
    \includegraphics[width=0.9\linewidth]{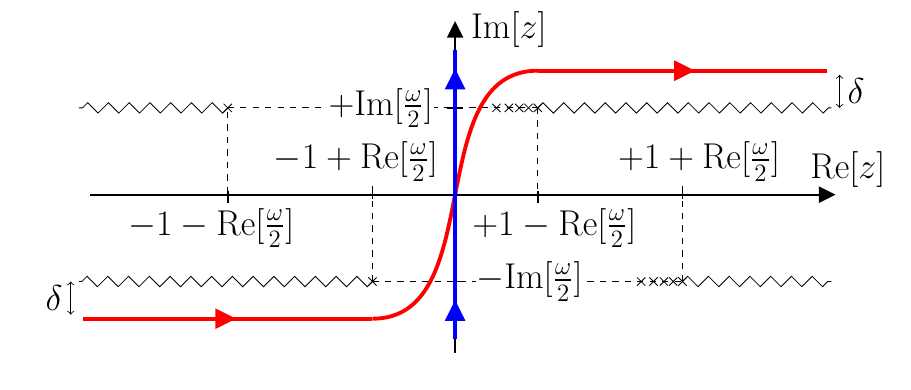}
    \caption{Integration contours for the outer loop $\omega$ (upper panel) and inner loop $z$ (lower panel), for the $1s$ reference state.
    The original integration path is displayed in red, while the rotated contour used in the computations is shown in blue.
    The lower contour is shown for the case of $\left(\text{Re}[\omega],\text{Im}[\omega]\right) > 0$.}
    \label{Fig:IntPath}
\end{figure}

The computation of Eq.~\eqref{Eq:AmpFinal} involves numerical integration over four radial coordinates, which are contained in $R_{J_1}$ and $R_{J_2}$, as well as the two loop energy variables, $\omega$ and $z$.
Carrying out the energy integrations along the real axis is problematic, so we rotate the contours in the complex $\omega$ and $z$ plane.
For the integral over $\omega$, we employ the same contour as in Ref.~\cite{yer2018} for the calculation of the self-energy, as shown in the upper panel of Fig.~\ref{Fig:IntPath}. For the integral over the energy in the fermion loop $z$, we make the plain Wick rotation to the imaginary axis, which is possible since $\text{Re}[\omega]$ is below the pair-production threshold, as shown in the lower panel of Fig.~\ref{Fig:IntPath}. Further details are presented in the End Matter.

One of the main challenges in the numerical evaluation arises from the multiple partial-wave expansions of the electron and photon propagators. After accounting for the angular-momentum selection rules, we are left with two infinite partial-wave sums and multiple finite summations. The two infinite sums must be truncated and extrapolated to infinity. The choice of the two expansion parameters is not unique. We choose them as (a) the Dirac angular-momentum quantum number $\kappa$ of the electron propagator in the self-energy loop and (b) the Dirac angular-momentum quantum number $\kappa'$ of one of the electron propagators in the closed fermion loop.

The convergence of the sum over the angular momentum in the fermion loop is relatively slow, so we have to include terms up to $\kappa'_{\rm max}=\pm16$ and estimate the tail by polynomial fitting in $1/|\kappa'|$, as in Ref.~\cite{syhc2026}. The sum over $\kappa$ converges much faster, allowing us to truncate the expansion typically at $\kappa_{\rm max}=\pm6$ and estimate the tail using a similar extrapolation procedure. In the high-$Z$ regime, the tail of the truncated expansion is very small, yielding a relatively high accuracy of about $0.5\%$ for uranium. Convergence becomes considerably slower, however, for lower nuclear charge numbers, and we achieve an accuracy of roughly $20\%$ for lithium, which is the lightest ion considered in this work.
%We estimate the uncertainty of the extrapolation as 200\% of the change in the final result caused by varying the truncation points of the partial-wave expansions and using different fitting functions to estimate the tail.

The large number of multipole expansion terms that need to be evaluated makes the computation highly demanding and necessitates extensive parallelization to keep the computational time manageable. We employ a hybrid parallelization scheme in which different integration points of the outer energy integral over $\omega$ are distributed among multiple processes. Each process spawns several threads that evaluate, in parallel, the radial integrals for all nonzero multipole combinations at the different integration points of the inner integral over $z$. Finally, the contributions from each multipole combination are combined with the corresponding Gaussian quadrature weights and then summed. A typical calculation takes about two days on 300 processor cores.

\begin{figure}
    \centering
    \includegraphics[width=1.\linewidth]{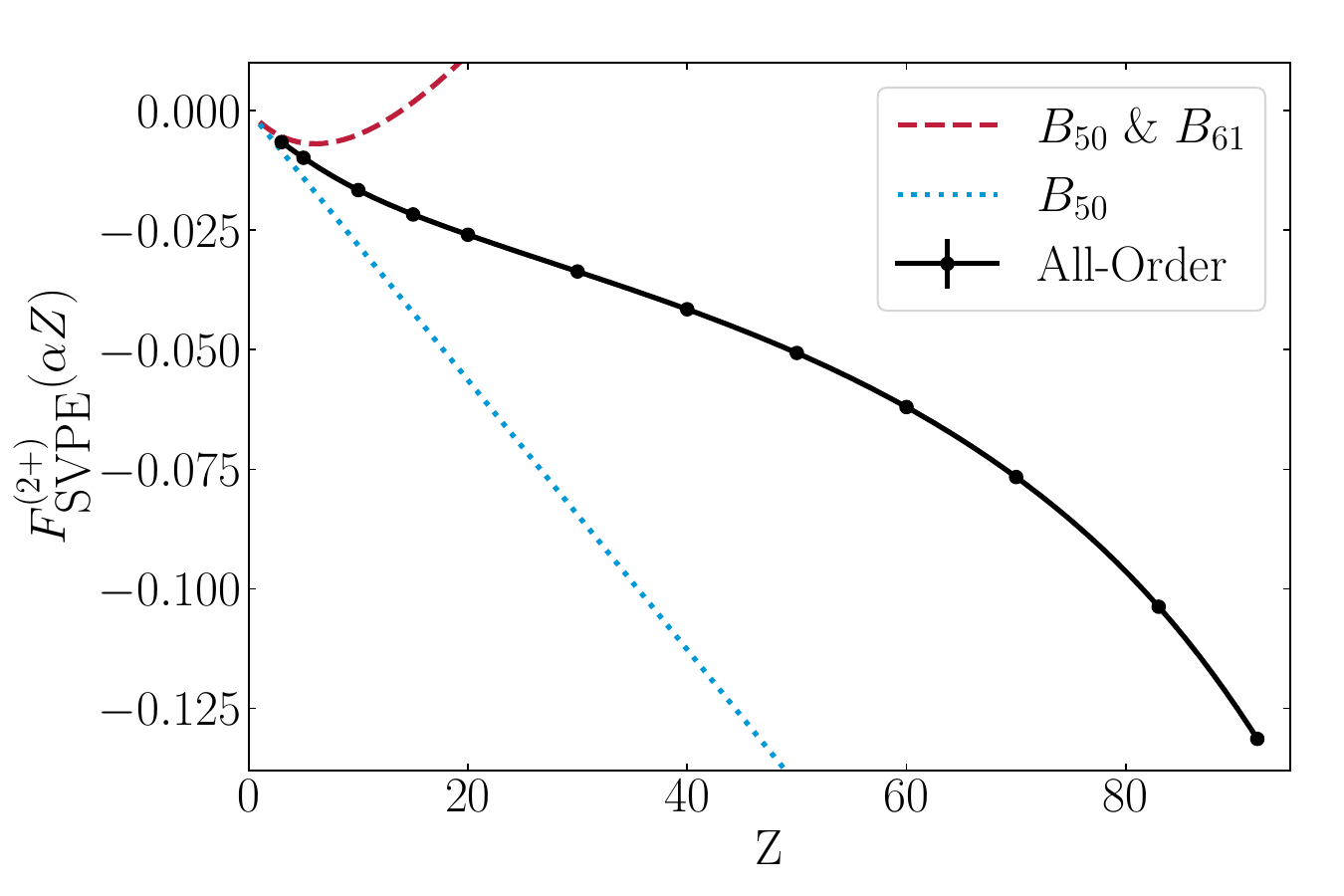}
    \caption{Higher-order SVPE contribution to the Lamb shift of the $1s$ state of hydrogen-like ions, in terms of the function $F_{\rm SVPE}$ defined in Eq.~\eqref{Eq:FaZ}.
    The all-order results (black dots) are compared with the contributions from the first two terms of the $\alpha Z$ expansion~\cite{pach1993,eides:94,dowling:10,sksi2019}, see Eq.~\eqref{Eq:FNRQED} (red dashed and blue dotted line).}
    \label{Fig:PlotZ}
\end{figure}

\begin{table}
\caption{The SVPE contributions to the Lamb shift of the $1s$ state of hydrogen-like ions, in terms
of $F_\text{SVPE}$ defined in Eq.~\eqref{Eq:FaZ}.
%Numerical uncertainties are given in parentheses.
} \label{Tab:FaZ}
\begin{ruledtabular}
\begin{tabular}{c D{.}{.}{2.4} D{.}{.}{2.8} D{.}{.}{2.8}}
%\\[-8pt]
 $Z$ & \multicolumn{1}{c}{$F_\text{SVPE}^{(0)}$} & \multicolumn{1}{c}{$F_\text{SVPE}^{(2+)}$} & \multicolumn{1}{c}{$F_\text{SVPE}$}\\[2pt]
\colrule
\\[-8pt]
3 & 0.1377 & -0.0066\,(13) & 0.1311\,(13) \\
5 & 0.1352 & -0.0099\,(13) & 0.1253\,(13) \\
10 & 0.1304 & -0.0166\,(8) & 0.1138\,(8) \\
15 & 0.1271 & -0.0217\,(4) & 0.1054\,(4) \\
20 & 0.1249 & -0.0260\,(4) & 0.0990\,(4) \\
30 & 0.1237 & -0.0336\,(4) & 0.0900\,(4) \\
40 & 0.1262 & -0.0416\,(4) & 0.0846\,(4) \\
50 & 0.1325 & -0.0507\,(4) & 0.0819\,(4) \\
60 & 0.1436 & -0.0620\,(3) & 0.0816\,(3) \\
70 & 0.1610 & -0.0766\,(3) & 0.0844\,(3) \\
83 & 0.1984 & -0.1037\,(3) & 0.0946\,(3) \\
92 & 0.2411 & -0.1314\,(3) & 0.1097\,(3)
\end{tabular}
\end{ruledtabular}
\end{table}

\textit{Numerical results---}
Numerical results for the SVPE correction are conveniently represented in terms of the dimensionless function $F_\text{SVPE}$ defined by
\begin{equation} \label{Eq:FaZ}
    E_\text{SVPE} = \left(\frac{\alpha}{\pi}\right)^2 \frac{(\alpha Z)^4}{n^3} F_\text{SVPE}(\alpha Z)~.
\end{equation}
Our numerical results for the SVPE correction obtained  for the $1s$ state of H-like ions are presented in Table~\ref{Tab:FaZ},
separated into the free-loop and higher-order contributions.
As seen from the table, the higher-order contribution is significant even for low-$Z$ ions, whereas for uranium it reduces the total value of the SVPE term by more than 50\%.
%Since calculations of the free-loop SVPE part are already well established \cite{yerokhin:08:twoloop},
%we concentrate our further analysis on the higher-order SVPE part.

The higher-order SVPE contribution has been previously investigated within the $\alpha Z$-expansion approach
\cite{pach1993,eides:94,dowling:10,sksi2019}. Its $\alpha Z$ expansion is given by
\begin{equation} \label{Eq:FNRQED}
\begin{aligned}
    F^{(2+)}_\text{SVPE}(\alpha Z) = &(\alpha Z) B_{50} + (\alpha Z)^2 \\
                                     &\times \Big[B_{61} \ln(\alpha Z)^{-2}+
     G_{\rm h.o.}^{(2+)}(\alpha Z)\Big] ~,
\end{aligned}
\end{equation}
where $B_{50} = -0.38615273$, $B_{61} = \frac{709\pi^2}{3456} - \frac{43}{36} = 0.830309...$,
and $G_{\rm h.o.}^{(2+)}$ is the remainder incorporating all higher orders in $\alpha Z$.
In Fig.~\ref{Fig:PlotZ}, we compare our all-order results for the higher-order SVPE correction with those obtained within the $\alpha Z$ expansion. As the figure shows, the all-order results approach the $\alpha Z$-expansion values at low nuclear charge numbers, providing an important verification of our numerical methods and the values of $B_{50}$ and $B_{61}$. However, higher-order terms in $\alpha Z$ make a substantial contribution to the total correction already at intermediate values of $Z$.

In the low-$Z$ region, the $Z\to0$ limiting value of the higher-order remainder $G_{\rm h.o.}^{(2+)}$ is
of importance for the hydrogen theory since it contributes to the theoretical uncertainty relevant for
determination of the Rydberg constant \cite{mohr:25:codata}. 
By extrapolating our all-order results for the higher-order remainder, we obtain the limiting value
$G_{\rm h.o.}^{(2+)}(Z = 0) =-2.7\,(10)$, see End Matter for details.
This contribution has not been included in Ref.~\cite{mohr:25:codata} and need to be added to 
the $B_{60}$ coefficient in Eq.~(46) of that work. 

\textit{Experimental consequences---}
We now examine the implications of our calculations for theoretical predictions of energy levels in HCIs.
The most immediate consequences are for H-like ions.
With the last missing contribution to the two-loop Lamb shift calculated in the present work, the theoretical precision of the ionization energies is substantially improved.
In order to illustrate this, Table~\ref{tab:hlike} summarizes the present status of the $1s$  Lamb shift in H-like uranium.
Most of the individual contributions are taken from the compilation of Ref.~\cite{yerokhin:15:Hlike},
with several recent updates. We added the higher-order two-loop vacuum-polarization correction
computed in Ref.~\cite{vyhk2025}
%now included into the VPVP entry,
and the all-order
result for the SVPE correction obtained in the present work. The finite nuclear size
correction is evaluated with the root-mean-square (rms) nuclear radius $r_C = 5.8571\,(33)$~fm
\cite{aam2013}, with inclusion of the nuclear deformation effect according to
Ref.~\cite{kozhedub:08}. The nuclear polarization correction was taken from
Ref.~\cite{nefiodov:96}, with a 20\% uncertainty ascribed by us here.

The total theoretical value contains two uncertainties, where the first one (``the'') is
the purely theoretical error, whereas the second uncertainty (``rms'') is
due to the experimental value of the nuclear radius. We observe that the theoretical uncertainty has been reduced by an order of magnitude
compared to the 2015 value \cite{yerokhin:15:Hlike}. The total
uncertainty of the theoretical prediction is now dominated by the uncertainty of
the nuclear radius.
The theoretical value agrees with the experimental result of Ref.~\cite{gumberidze:05:prl},
but the experimental accuracy is not yet sufficient to probe the two-loop QED effects.

\begin{table}
\caption{Lamb shift of the $1s$ state of $^{238}$U$^{91+}$, in eV.
TW stands for ``this work''. To obtain the ionization energy, the point-nucleus Dirac energy should be added to the Lamb shift,
$E_D = mc^2[\sqrt{1-(\alpha Z)^2}-1]$.} \label{tab:hlike}
\begin{ruledtabular}
\begin{tabular}{ll D{.}{.}{4.12}r}
\multicolumn{2}{l}{ Term} & \multicolumn{1}{c}{Value \ \ \ \ \ \ \ } & Ref.\\
 \colrule
\\[-7pt]
One-loop QED        &SE  & 355.06   & \cite{mohr:74:a,yerokhin:15:Hlike} \\
                    &VP  & -88.61   & \cite{soff:88:vp,yerokhin:15:Hlike} \\
Two-loop QED        &SESE&  -1.57   & \cite{yerokhin:03:prl,yerokhin:25:sese} \\
                    &SEVP&   1.13   & \cite{yerokhin:08:twoloop} \\
                    &SVPE&   0.06   & TW \\
                    &VPVP&  -0.89   & \cite{yerokhin:08:twoloop,vyhk2025} \\
                    &FNS &   0.02\,(2) & TW \\
Three-loop QED      &&  \pm0.01 & TW \\
Nuclear size                 && 198.51\,(19)_{\rm rms} & \cite{kozhedub:08,yerokhin:15:Hlike} \\
\multicolumn{2}{l}{
Nuclear recoil        }&   0.47   & \cite{artemyev:95:pra,yerokhin:15:Hlike} \\
\multicolumn{2}{l}{
Nuclear polarization  }&  -0.21\,(4) & \cite{nefiodov:96} \\
\colrule
Theory              && 463.97\,(5)_{\rm the}(19)_{\rm rms} \\
Previous theory     && 463.84\,(49)_{\rm the}(19)_{\rm rms} & \cite{yerokhin:15:Hlike}\\
Experiment          && 460.\,(5)        & \cite{gumberidze:05:prl} \\
\end{tabular}
\end{ruledtabular}
\end{table}

A comparison between theory and experiment is much more advantageous for Li-like ions, where substantially higher experimental accuracy has been achieved. In particular, Ref.~\cite{beiersdorfer:98} reported a measurement of the $2p_{3/2}$–$2s$ transition energy in Li-like bismuth with a precision of $0.04$~eV, two orders of magnitude better than that achieved for H-like ions. Although our calculations have been performed only for the $1s$ state, the strong correlation between QED corrections for different $ns$ states allows us to obtain a reliable estimate of the SVPE correction for the $2p_{3/2}$–$2s$ transition energy.
Specifically,
we examine the ratio of QED corrections for the $1s$ energy and the $2s$-$2p_{3/2}$ energy difference
in H-like bismuth.
This ratio is  $6.6$ for the one-loop self-energy correction, $6.1$ for the two-loop self-energy,
and $5.9$ for the free-loop SVPE correction \cite{yerokhin:15:Hlike}.
We therefore adopt $5.9 \pm 50\%$ as a conservative estimate of the corresponding ratio for the higher-order SVPE correction.

\begin{table}
\caption{The $2p_{3/2} $--$ 2s$ transition energy in $^{209}$Bi$^{80+}$, in \si{\eV}. } \label{tab:lilike}
\begin{ruledtabular}
\begin{tabular}{lc D{.}{.}{4.16}r}
  \multicolumn{2}{c}{Term} & \multicolumn{1}{c}{Value} & \multicolumn{1}{c}{Ref.}\\
\colrule
\\[-7pt]
Electron structure             & & 2814.393\,(3)_{\rm the}(9)_{\rm rms} & \cite{yhk2025b},TW \\
One-loop QED          & & -27.486\,(1)  & \cite{yerokhin:15:Hlike} \\
\multicolumn{2}{l}{
Screened one-loop QED}  &   1.140\,(3)  & \cite{yhk2025b} \\
Two-loop QED     & SESE &   0.141   & \cite{yerokhin:03:prl,yerokhin:25:sese} \\
                 & SEVP &  -0.095   & \cite{yerokhin:08:twoloop} \\
                 & SVPE &  -0.006\,(3)   & TW \\
                 & VPVP &   0.080   & \cite{yerokhin:08:twoloop,vyhk2025} \\
\multicolumn{2}{l}{
Screened two-loop QED}  &  -0.007\,(5) & \cite{yhk2025b} \\
Nuclear recoil        & &  -0.062\,(2) & \cite{yhk2025b} \\
\multicolumn{2}{l}{
Nuclear polarization  } &   0.005\,(2) & TW \\
\colrule
Theory                & & 2788.102\,(8)_{\rm the}(9)_{\rm rms} \\
Previous theory       & & 2788.116\,(35)_{\rm the}(9)_{\rm rms}(24)_{\rm mod}    & \cite{yhk2025b} \\
                      & & 2788.04\,(5)               & \cite{sapirstein:11} \\
Experiment            & & 2788.139\,(39)       & \cite{beiersdorfer:98} \\
\end{tabular}
\end{ruledtabular}
\end{table}

Table~\ref{tab:lilike} summarizes theoretical contributions to the  $2p_{3/2}$-$2s$ transition energy in
Li-like bismuth.
Most of the individual contributions correspond to Ref.~\cite{yhk2025b},
with several updates. First, we updated the treatment of the finite nuclear size effect to eliminate
the model dependence of the nuclear charge distribution. This was achieved by
using the electron-nucleus scattering data to constrain
the three first ratios of moments of the nuclear charge distribution, following the procedure described in
Ref.~\cite{yerokhin:26:rms}. Specifically, we used the rms radius $r_C = 5.5211\,(26)$~fm
\cite{aam2013} and the moment ratios $V_{24} \equiv r_C/\langle r^4\rangle^{1/4}=  0.9409\,(2)$
and $V_{26} \equiv r_C/\langle r^6\rangle^{1/6} =  0.8990\,(8)$ from Ref.~\cite{yerokhin:26:rms}
in our calculations.
It was demonstrated in Ref.~\cite{yerokhin:26:rms} that the remaining nuclear model dependence
is less than $0.001$~eV and is completely negligible at present.
Furthermore, we included the higher-order VPVP correction
computed in Ref.~\cite{vyhk2025},  and the higher-order
SVPE correction obtained in the present work.
In addition, we estimated the nuclear polarization correction, by utilizing the
approach of Ref.~\cite{nefiodov:96} with the implementation described in Ref.~\cite{viatkina:23}.

As demonstrated in Table~\ref{tab:lilike},  we have reduced the theoretical uncertainty of the $2p_{3/2}$–$2s$ transition energy in Li-like bismuth by a factor of five. The theoretical prediction is now 3.2 times more precise than the experimental value, highlighting the need for further improvements in experimental accuracy. The largest contribution to the theoretical uncertainty currently arises from the screening of two-loop QED effects, but several other contributions have comparable uncertainties.

\textit{Conclusion---}
We reported a calculation of the last unknown contribution to the two-loop Lamb shift, induced by
the electron self-energy diagram
with the vacuum-polarization insertion into the photon line. The calculation was carried out to
all orders in the nuclear binding strength parameter $\alpha Z$, for the $1s$ state of
hydrogen-like ions for a wide range of nuclear charge numbers, $Z = 3$ -- $92$. In the low-$Z$
region, our all-order results approach known results obtained within the expansion in $\alpha Z$. 
By extrapolating our results to $Z = 0$, we determined the previously unknown contribution to the 
so-called $B_{60}$ coefficient, which is needed to extract the Rydberg constant from hydrogen spectroscopy data.

Our calculation brings to completion a decades-long project of evaluating the two-loop Lamb shift to all orders in $\alpha Z$. It improves the accuracy of theoretical predictions for energy levels and transition energies in heavy hydrogen-like and lithium-like ions, enabling more stringent comparisons between theory and available as well as future experimental results.
These results remove an important theoretical limitation in the description of the two-loop Lamb shift and pave the way for more precise tests of quantum electrodynamics in the strong-field regime using highly charged ions, as well as for accurate determinations of nuclear charge radii.

%%%%%%%%%%%%%%%%%%%%%%%%%%%%%%%%%%%%%%%%%%%%%%%%%%%%%%%%%%%%%%%%%%%%%%%%%%%%%%%%%%%%%%%%%%%

\bibliography{./SVPE.references}

\clearpage
\appendix
\section*{End Matter}
% Reset and rename counters
\setcounter{equation}{0}
\setcounter{figure}{0}
\setcounter{table}{0}
\setcounter{page}{1}
\subsection{Computational Details}
\begin{figure}[hb]
    \centering
    \includegraphics[width=1.\linewidth]{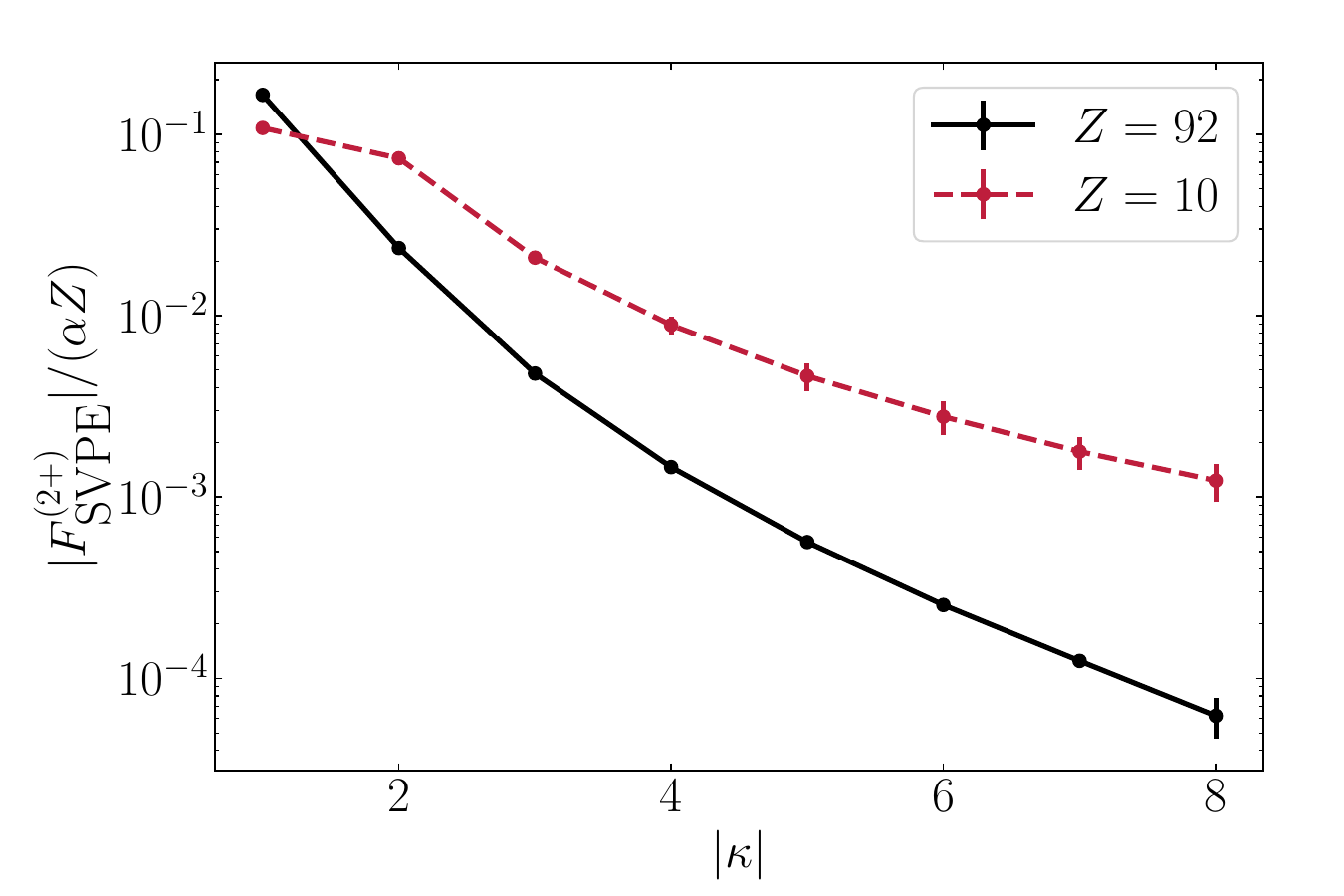}
    \caption{Contribution of different multipoles $\kappa$ of the electron propagator outside the closed Fermion loop to the SVPE diagram for the $1s_{1/2}$ state of hydrogen-like ions divided by $\left(\frac{\alpha}{\pi}\right)^2 (\alpha Z)^5$. We compare hydrogen-like uranium (black solid line) to hydrogen-like neon (red dashed line).}
    \label{Fig:Plotk}
\end{figure}
To accelerate the convergence of the numerical integrals, we use the same technique as in Ref.~\cite{syhc2026} and explicitly subtract the spurious non-gauge-invariant terms, which are obtained by the replacement in the closed Fermion loop
\begin{equation} \label{Eq:RepRen}
\begin{aligned}
    G_{23}&\left(z+\frac{\omega}{2}\right) G_{32}\left(z-\frac{\omega}{2}\right) \\
    &\to \frac{1}{2}\frac{d^2}{dz^2}\left[ V_2F_{23}\left(z+\frac{\omega}{2}\right)F_{32}\left(z-\frac{\omega}{2}\right)V_3\right]~,
\end{aligned}
\end{equation}
where $G_{ij}(z) =  G\left(\boldsymbol{x}_i, \boldsymbol{x}_j, z\right)$ and  $F_{ij}(z) =  G\left(\boldsymbol{x}_i, \boldsymbol{x}_j, z\right)\vert_{Z=0}$ are the Dirac-Coulomb and free Dirac Green's functions and $V_i = V(x_i)$ is the Coulomb potential.

The radial and loop-energy integrals in $E_\text{SVPE}$ are numerically evaluated using Gaussian quadrature with the density of integration points chosen large enough so that the relative uncertainty in the final result is less than $10^{-5}$ in the high-$Z$ regime. For low $Z$, there are large numerical cancellations that can reduce the accuracy. The integration in the complex $\omega$ plane is performed along the intervals $]\Delta-i\infty,\Delta]$, $[\Delta,0]$, $[0,\Delta]$ and $[\Delta,\Delta+i\infty[$, where $\Delta = \alpha Z \epsilon_a$ and the appropriate branch of the photon propagators has to be chosen for each interval.

The largest uncertainty in our computation originates from the multiple expansions of the virtual particles. As described in the main text, we truncate the sum over the angular momentum inside the fermion loop at $\kappa'_{\rm max} = \pm 16$ and estimate the tail using the same extrapolation technique as in Ref.~\cite{syhc2026}. The sum over $\kappa$ converges much faster, and therefore we only sum up to $\kappa_{\rm max} = \pm 6$ in our calculation and extrapolate from there. For $Z = 92$ and $Z=10$, we have performed calculations up to $\kappa_{\rm max} = \pm 8$ to estimate the precision of this truncation method. In the high-$Z$ regime, the remaining tail is very small, whereas convergence is much slower for low nuclear charge numbers, see Fig.~\ref{Fig:Plotk}. We estimate the uncertainty as 200\% of the influence on the final energy shift when varying $\kappa_{\rm max}$ and changing the fitted function used to estimate the tail of the multipole expansions. Overall, we achieve a relative accuracy of around 0.5\% for the $1s_{1/2}$ state of uranium. In the low-$Z$ regime, the uncertainty is significantly higher due to large numerical cancellation errors and a much slower convergence of the multipole expansions.

\subsection{Extrapolation to Small Charge Numbers}
\begin{figure}[hb]
    \centering
    \includegraphics[width=1.\linewidth]{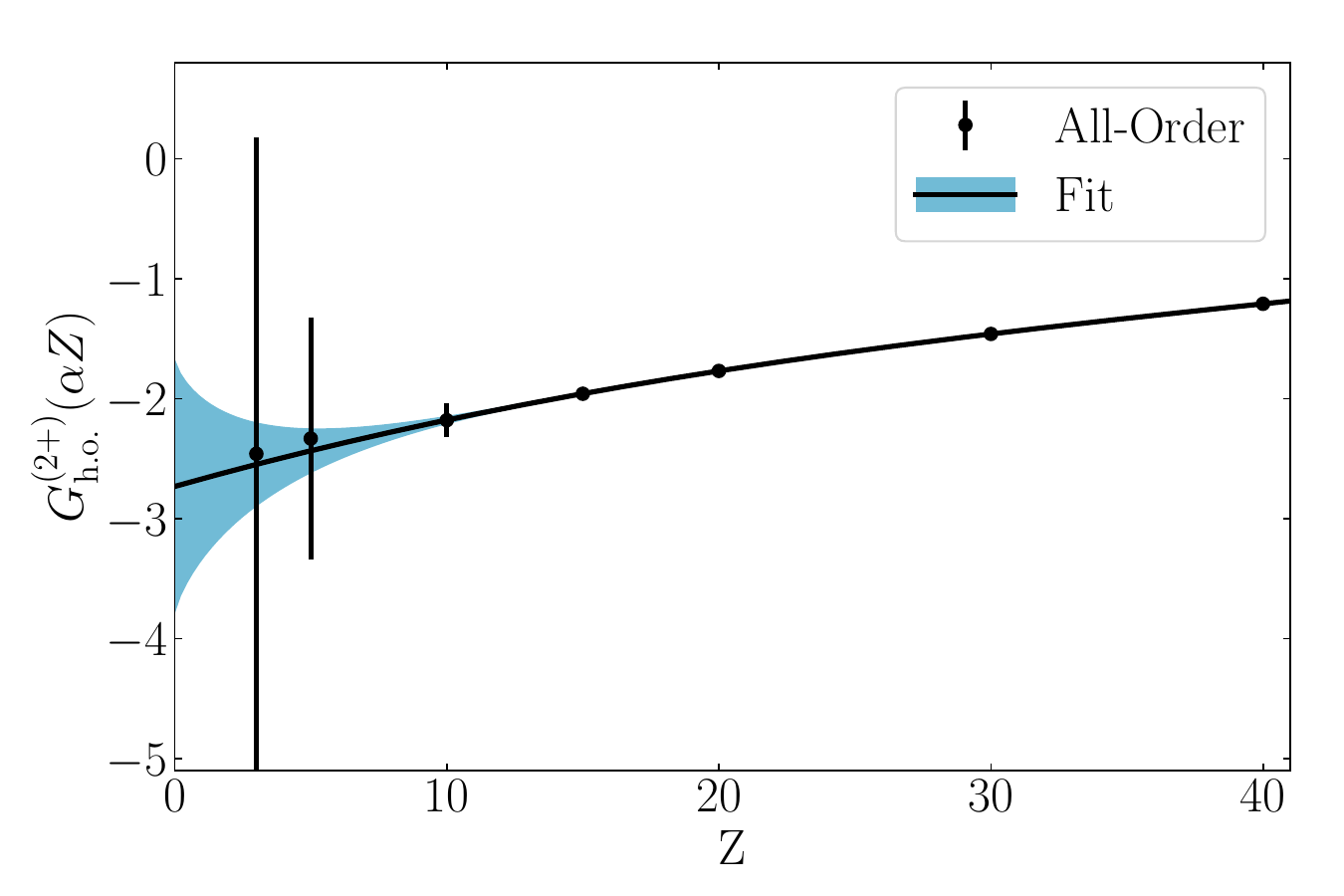}
    \caption{Higher-order remainder $G^{(2+)}_\text{h.o.}$ as a function of $Z$.
     Black dots represent the numerical all-order data, black solid line is the best fit, the blue shaded area shows the uncertainty.
}
    \label{Fig:PlotZ0}
\end{figure}

We extrapolate our all-order results for the higher-order remainder $G_{\rm h.o.}^{(2+)}$, defined in Eq.~(\ref{Eq:FNRQED}) of the main text, to the low-$Z$ region. The extrapolation is complicated by the presence of unknown logarithmic terms in its expansion in $\alpha Z$. These terms exhibit a rapidly-varying structure only at very small $Z$, where the numerical convergence of our calculations is poor, making their contributions difficult to determine reliably by fitting. We therefore fit our results with polynomials and separately estimate an additional uncertainty associated with the unknown logarithmic terms.
Specifically, we fit our all-order data for $Z\leq 40$, weighted by their numerical uncertainties, with polynomials in $\alpha Z$ of different orders. We estimate the uncertainty from the spread of the extrapolated values obtained with different polynomials and also account for the shift induced by including a single logarithmic term with a coefficient $B_{71}=\pm 10$.
In the result, we obtain 
\begin{equation}
    G_{\rm h.o.}^{(2+)}(Z = 0) = -2.7\,(10)~,
\end{equation}
with results graphically represented in Fig.~\ref{Fig:PlotZ0}.

\end{document}